\documentclass[11pt]{article}

\usepackage[utf8]{inputenc}
\usepackage[T1]{fontenc}
\usepackage{lmodern}
\usepackage[margin=1in]{geometry}
\usepackage{amsmath,amssymb}
\usepackage{graphicx}
\usepackage{booktabs}
\usepackage{caption}
\usepackage[hidelinks]{hyperref}

\graphicspath{{figures/}}

\title{\textbf{HistoFID}\\[2pt]
\large Calibrating Fr\'echet-distance evaluation across pathology foundation models}
\date{}

\author{
Swapnil Bhat$^{1}$, Devansh Lalwani$^{1,\ast}$, Maulik Shah$^{1}$, Sheena Alphones$^{2}$,\\
Rimlee Dutta$^{3}$, Nikita Mulchandani$^{4}$, Roshani Gala$^{4}$, Jay Mehta$^{4}$\\[6pt]
{\small $^{1}$Turocrates.ai, Mumbai, India \quad $^{2}$NM Medical, Mumbai, India \quad $^{4}$Neuberg OncoPath, Mumbai, India}\\[1pt]
{\small $^{3}$All India Institute of Medical Sciences (AIIMS), New Delhi, India}\\[1pt]
{\small $^{\ast}$Corresponding author: \texttt{founders@turocrates.ai}}
}

\begin{document}
\maketitle

\begin{abstract}
\noindent
The Fr\'echet Inception Distance (FID) compares two image sets by fitting a Gaussian to the
features of a fixed network and measuring the distance between the two Gaussians. In digital
pathology the Inception network is routinely replaced by a histology foundation model, on the
assumption that a domain encoder gives a more meaningful score. We show that this choice changes
the result. For one fixed pair of tile sets, the raw Fr\'echet distance varies by about thirty-fold
across six common encoders, and the ordering does not follow embedding dimension, so a raw score
cannot be read without naming the encoder. Using a held-out in-house cohort (about 500,000 H\&E and
immunohistochemistry tiles from 2,119 slides) and a public TCGA-BRCA cohort (100 slides), we
benchmark Inception-v3, Phikon-v2, CONCH, UNI2-h, Virchow2 and Prov-GigaPath across within-cohort
baselines, cross-cohort drift, controlled perturbations, image compression, stain normalization,
and two generative models (PixCell-256 and CytoSyn). Expressing each distance as a ratio to the
encoder's own within-cohort floor restores comparability, cutting the across-encoder coefficient of
variation by about 89\% within cohort and about 58\% across cohorts. The encoders then separate into
a sensitive group (CONCH, Phikon-v2, Inception-v3) and an invariant group (UNI2-h, Virchow2,
Prov-GigaPath), and the same split decides which generative model is judged more realistic, so the
encoder can change the conclusion of a generative evaluation. At the slide level, an attention-pooling
encoder registers per-slide composition that a pooled patch distance cannot see, raising the distance
by about 320-fold on matched cohorts. Using the same protocol we evaluate TuroCompress, a proprietary
pathology codec, which reaches the highest reconstruction fidelity at the smallest file size among the
codecs tested. We release the normalization protocol, the per-encoder perturbation panel and the
feature extracts.
\end{abstract}

\section{Introduction}

The Fr\'echet Inception Distance compares two image sets by fitting a Gaussian to the activations of a
fixed Inception-v3 network and measuring the Fr\'echet distance between the two Gaussians
\cite{heusel2017,szegedy2016}. It is the default summary statistic for generative image evaluation and
is increasingly used in digital pathology to compare synthetic and real tiles, to quantify batch or
scanner effects, and to assess image compression. In pathology the Inception network is usually
replaced by a foundation model trained on histology \cite{chen2024uni,vorontsov2024virchow,zimmermann2024virchow2,filiot2023,filiot2024phikon,lu2024conch,xu2024gigapath},
on the assumption that a domain encoder yields a more meaningful feature space.

This substitution is rarely examined. Because the distance is computed entirely in the encoder's
feature space, its value inherits that space's scale, geometry and learned invariances, and different
encoders differ in all three. This paper quantifies how much the reported distance depends on the
encoder for a fixed comparison, and shows that the dependence can be removed without discarding the
metric.

\begin{figure}[t]
\centering
\includegraphics[width=\textwidth]{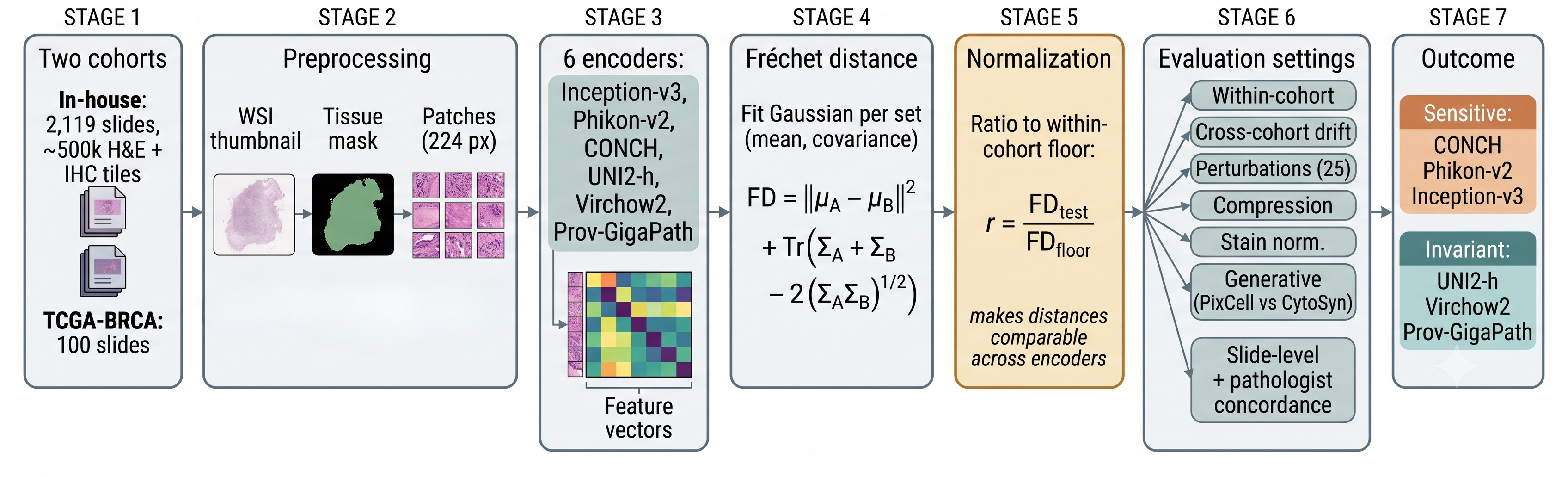}
\caption{Study workflow: two cohorts, six feature extractors, Fr\'echet distance with sample-size
matching, ratio normalization, and five evaluation settings plus a generative comparison.}
\label{fig:workflow}
\end{figure}

This paper makes three contributions.
\begin{enumerate}
  \item We quantify the dependence of the Fr\'echet distance on the encoder across six models and two
  cohorts, and show that the raw distance for an identical comparison spans about thirty-fold and does
  not order by embedding dimension.
  \item We give a simple normalization, the ratio of a measured distance to the encoder's own
  within-cohort floor, that makes distances comparable across encoders, and we validate it on
  independent comparisons.
  \item We characterize how encoders differ in sensitivity to distribution shift, and show that this
  difference changes the outcome of cohort, stain and generative evaluations.
\end{enumerate}
Figure~\ref{fig:workflow} summarizes the study.

\section{Background and related work}

The Fr\'echet Inception Distance \cite{heusel2017} is biased for finite samples, and the bias depends
on the model being evaluated, so two models can be ranked differently for reasons unrelated to image
quality \cite{chong2020}. The bias falls with sample size; we control for it by matching sample sizes
between a measured comparison and its reference. Sample-based precision and recall \cite{kynkaanniemi2019}
and the Feature Likelihood Divergence \cite{jiralerspong2023} separate fidelity from coverage, and we
report both alongside the distance.

Histology encoders used here include UNI2-h \cite{chen2024uni}, Virchow2
\cite{vorontsov2024virchow,zimmermann2024virchow2}, Phikon-v2 \cite{filiot2023,filiot2024phikon}, the
vision-language model CONCH \cite{lu2024conch}, and Prov-GigaPath \cite{xu2024gigapath}. Most are
trained with self-supervision, Phikon-v2 and the others with DINOv2 \cite{oquab2024dinov2}, on cohorts
ranging from a single public archive to billions of proprietary tiles. PixCell-256 is a diffusion model
conditioned on UNI embeddings \cite{yellapragada2025pixcell}; CytoSyn is a flow-based model with a
scalable interpolant transformer backbone \cite{ma2024sit,leng2025repae}. We use both as test cases for
encoder-dependent generative evaluation.

\section{Materials and methods}

\subsection{Cohorts}

\textbf{In-house cohort.} 2,119 whole-slide images of breast pathology (H\&E and immunohistochemical
stains including ER, PR, KI67 and HER2), supplied as pre-extracted $512\times512$ tiles in HDF5, about
500,000 tiles in total. The archive's train, validation and test split defines slide-disjoint reference
and held-out sets.

\textbf{TCGA-BRCA cohort.} 100 diagnostic slides obtained through the Genomic Data Commons
\cite{grossman2016}. Tissue was detected by Otsu thresholding on a slide thumbnail with morphological
cleanup, and tiles were extracted at 224 pixels and 0.5 micrometres per pixel, then partitioned
slide-disjointly. We keep this cohort smaller than the in-house cohort on purpose. TCGA is the most
heavily reused public pathology archive, and it overlaps the training data of some encoders evaluated
here, most clearly Phikon-v2, which was trained on public cohorts that include TCGA. Using TCGA as the
primary reference would therefore evaluate those encoders partly on their own training distribution. We
instead use the held-out in-house cohort as the primary reference and treat TCGA as a secondary public
comparison.

\subsection{Feature extractors}

\begin{table}[t]
\centering
\caption{Patch-level feature extractors used to compute the Fr\'echet distance.}
{\small
\begin{tabular}{@{}llll@{}}
\toprule
Encoder & Dim & Objective / type & Training cohort \\
\midrule
Inception-v3 & 2048 & ImageNet supervised & ImageNet \\
Phikon-v2 & 1024 & DINOv2 & $>$100 public cohorts, 460M tiles (incl.\ TCGA) \\
CONCH & 512 & CLIP (vision encoder) & 1.17M image-caption pairs \\
UNI2-h & 1536 & DINOv2 & BWH/MGH, $\sim$100M tiles \\
Virchow2 & 1280 & DINOv2 (mixed magnification) & 3.1M proprietary slides \\
Prov-GigaPath & 1536 & DINOv2 (tile encoder) & 1.3B tiles, Providence \\
\bottomrule
\end{tabular}}
\label{tab:encoders}
\end{table}

Six patch encoders were evaluated (Table~\ref{tab:encoders}), with Inception-v3 \cite{szegedy2016} as
the natural-image reference. Tiles were preprocessed identically per encoder (resize, center crop,
encoder-specific normalization) and features were stored in float16.

\subsection{Fr\'echet distance and normalization}

For feature sets $X$ and $Y$ with means $\mu_X,\mu_Y$ and covariances $\Sigma_X,\Sigma_Y$, the
Fr\'echet distance is the closed-form squared 2-Wasserstein distance between the fitted Gaussians
\cite{dowson1982,heusel2017},
\begin{equation}
\mathrm{FD}(X,Y) \;=\; \lVert \mu_X - \mu_Y \rVert_2^{2}
\;+\; \operatorname{Tr}\!\Big( \Sigma_X + \Sigma_Y - 2\big( \Sigma_X \Sigma_Y \big)^{1/2} \Big).
\label{eq:fd}
\end{equation}
The matrix square root was computed in double precision and its imaginary residual required to be below
$10^{-3}$. Because the sample mean and covariance are estimated from finitely many features, the
estimator is biased upward, and the excess decays with the per-side sample count $N$,
\begin{equation}
\mathbb{E}\!\left[\widehat{\mathrm{FD}}_N\right] \;=\; \mathrm{FD}_{\infty} \;+\; \frac{c}{N}
\;+\; o\!\left(\tfrac{1}{N}\right),
\label{eq:bias}
\end{equation}
where $\mathrm{FD}_\infty$ is the infinite-sample limit and the coefficient $c$ depends on the feature
covariance and therefore on the encoder \cite{chong2020}. Because $c$ is encoder-dependent, matching
$N$ does not equalize the bias across encoders; it removes sample size as a confound so that every
comparison in this paper is evaluated at the same $N$ (25,000 per side unless noted).

To compare distances across encoders we divide each measured distance by the encoder's own
within-cohort floor. The floor is the distance between two disjoint reference halves $R^{A},R^{B}$ of
the same cohort at matched size, so it captures the value produced by finite-sample estimation alone,
\begin{equation}
\mathrm{FD}^{\mathrm{floor}}_{e} \;=\; \mathrm{FD}_e\!\big( \Phi_e(R^{A}),\, \Phi_e(R^{B}) \big),
\qquad R^{A}\cap R^{B}=\varnothing,\;\; \lvert R^{A}\rvert=\lvert R^{B}\rvert=N,
\label{eq:floor}
\end{equation}
where $\Phi_e$ is encoder $e$'s embedding map. The reported quantity is the ratio and its logarithm,
\begin{equation}
r_e \;=\; \frac{\mathrm{FD}^{\mathrm{test}}_{e}}{\mathrm{FD}^{\mathrm{floor}}_{e}},
\qquad
\log r_e \;=\; \log \mathrm{FD}^{\mathrm{test}}_{e} - \log \mathrm{FD}^{\mathrm{floor}}_{e}.
\label{eq:ratio}
\end{equation}
Both the test distance and the floor carry the same per-encoder feature scale, so the ratio cancels it
and yields a number that is nearly encoder-invariant, whereas a difference would leave that scale intact.
The ratio is a ratio of squared distances. To quantify how much the encoders agree on one comparison we
use the coefficient of variation of the six per-encoder values $F_1,\dots,F_n$,
\begin{equation}
\mathrm{CV} \;=\; \frac{\sigma}{\mu} \;=\; \frac{1}{\mu}\sqrt{\frac{1}{n}\sum_{i=1}^{n}\bigl(F_i-\mu\bigr)^{2}},
\qquad \mu \;=\; \frac{1}{n}\sum_{i=1}^{n} F_i,
\label{eq:cv}
\end{equation}
which is dimensionless; a small CV of the normalized values indicates the encoders assign concordant
ratios.

\subsection{Evaluation settings}

The settings were: within-cohort baseline; cross-cohort drift (in-house versus TCGA); a panel of 25
controlled perturbations applied to 1,000 tiles, normalized within each encoder so that sensitivities
are comparable across encoders; image compression (JPEG, JPEG 2000, JPEG XL, and TuroCompress, a
proprietary compression codec from Turocrates.ai) scored by distance and by SSIM \cite{wang2004} and
LPIPS \cite{zhang2018}; stain normalization (Macenko \cite{macenko2009}, Reinhard \cite{reinhard2001},
Vahadane \cite{vahadane2016}); and a generative comparison of PixCell-256 \cite{yellapragada2025pixcell}
and CytoSyn scored by distance, precision and recall \cite{kynkaanniemi2019}, and FLD
\cite{jiralerspong2023}.

\subsection{Evaluation metrics}

\textbf{Precision and recall.} We use the improved, manifold-based precision and recall
\cite{kynkaanniemi2019}. Each feature set is approximated by a union of hyperspheres, one per sample with
radius equal to the distance to that sample's $k$-th nearest neighbor, and membership is the indicator
\begin{equation}
f(\phi,\Phi)=
\begin{cases}
1, & \text{if } \lVert \phi - \phi' \rVert_2 \le \lVert \phi' - \mathrm{NN}_k(\phi',\Phi) \rVert_2
\text{ for at least one } \phi' \in \Phi,\\
0, & \text{otherwise.}
\end{cases}
\label{eq:indicator}
\end{equation}
Precision is the fraction of generated features inside the real manifold and reads as fidelity; recall
is the fraction of real features inside the generated manifold and reads as coverage,
\begin{equation}
\mathrm{precision}(\Phi_r,\Phi_g) = \frac{1}{\lvert \Phi_g \rvert}\!\!\sum_{\phi_g \in \Phi_g}\!\! f(\phi_g,\Phi_r),
\qquad
\mathrm{recall}(\Phi_r,\Phi_g) = \frac{1}{\lvert \Phi_r \rvert}\!\!\sum_{\phi_r \in \Phi_r}\!\! f(\phi_r,\Phi_g).
\label{eq:pr}
\end{equation}

\textbf{Feature Likelihood Divergence.} FLD \cite{jiralerspong2023} fits a Gaussian kernel density with a
per-sample bandwidth on the generated features and scores the held-out real features under it, offset so
an ideal generator scores near zero,
\begin{equation}
p_{\boldsymbol{\sigma}}\!\left(\mathbf{x}\mid\mathcal{D}_{\mathrm{gen}}\right)
=\frac{1}{m}\sum_{j=1}^{m}\mathcal{N}\!\left(\varphi(\mathbf{x});\,\varphi(\mathbf{x}^{\mathrm{gen}}_{j}),\,\sigma_{j}^{2}\mathbf{I}_{d}\right),
\quad
\mathrm{FLD} = -\frac{100}{d}\,\frac{1}{n}\sum_{i=1}^{n}\log p_{\hat{\boldsymbol{\sigma}}}\!\left(\mathbf{x}^{\mathrm{test}}_{i}\mid\mathcal{D}_{\mathrm{gen}}\right) - C,
\label{eq:fld}
\end{equation}
where the bandwidths $\hat{\boldsymbol{\sigma}}$ are fit by maximum likelihood on a disjoint real train
split. Because the bandwidths are tuned on training data, a generated point that copies a training
example contributes little to the held-out likelihood, so FLD penalizes memorization; it is
lower-is-better and near zero for an ideal generator, and can dip slightly below zero.

\textbf{SSIM and LPIPS.} For compression we use two full-reference pixel metrics. SSIM \cite{wang2004}
compares local luminance, contrast and structure,
\begin{equation}
\mathrm{SSIM}(x,y) = \frac{\left(2\mu_x\mu_y + C_1\right)\left(2\sigma_{xy} + C_2\right)}
{\left(\mu_x^2 + \mu_y^2 + C_1\right)\left(\sigma_x^2 + \sigma_y^2 + C_2\right)},
\qquad C_1=(K_1 L)^2,\; C_2=(K_2 L)^2,
\label{eq:ssim}
\end{equation}
with $K_1=0.01$, $K_2=0.03$ and dynamic range $L$; it lies in $[-1,1]$, is $1$ for identical images, and
is reported as the mean over local windows. LPIPS \cite{zhang2018} is a learned perceptual distance,
the weighted, channel-normalized feature difference across layers of a fixed network,
\begin{equation}
d_{\mathrm{LPIPS}}(x, x_0) = \sum_{l} \frac{1}{H_l W_l} \sum_{h,w}
\bigl\lVert w_l \odot \bigl( \hat{y}^{\,l}_{hw} - \hat{y}^{\,l}_{0,hw} \bigr) \bigr\rVert_2^2,
\qquad \hat{y}^{\,l}_{hw} = \frac{F^{\,l}_{hw}(x)}{\lVert F^{\,l}_{hw}(x) \rVert_2},
\label{eq:lpips}
\end{equation}
where $w_l$ are the learned non-negative channel weights; it is a distance, so smaller is better, and
complements SSIM.

\textbf{Invariance and perturbation sensitivity.} The stability of an encoder $f_i$ under a $90$-degree
rotation $R_{90}$ is the cosine between a tile's feature vector and that of its rotated copy,
\begin{equation}
c_i \;=\; \cos\!\big(f_i(t),\, f_i(R_{90}\,t)\big)
\;=\; \frac{\langle f_i(t),\, f_i(R_{90}\,t)\rangle}{\lVert f_i(t)\rVert_2\,\lVert f_i(R_{90}\,t)\rVert_2},
\label{eq:cosine}
\end{equation}
with $1$ meaning the two embeddings point in the same direction ($90$-degree-rotation stability). For the
perturbation panel, the raw disruption $d_{ij}$ is the Fr\'echet distance between the clean and
perturbed feature sets of encoder $i$ under perturbation $j$, and each encoder's row is min-max
normalized so its least and most disruptive perturbations map to $0$ and $1$,
\begin{equation}
s_{ij} \;=\; \frac{d_{ij} - \min_{j'} d_{ij'}}{\max_{j'} d_{ij'} - \min_{j'} d_{ij'}}.
\label{eq:sens}
\end{equation}
A value of $1$ marks that encoder's most disruptive perturbation only and is not comparable in absolute
terms across encoders.

\textbf{Concordance.} To relate feature distance to expert perception we use the Spearman rank
correlation between the reader's similarity ordering of image variants and the encoder-distance ordering
of the same variants, defined as the Pearson correlation of the ranks,
\begin{equation}
\rho_{s} \;=\; \frac{\operatorname{cov}\!\left(R(X),\,R(Y)\right)}{\sigma_{R(X)}\,\sigma_{R(Y)}}
\;=\; 1 - \frac{6\sum_{i=1}^{n} d_i^{2}}{n\left(n^{2}-1\right)} \;\; (\text{distinct ranks}),
\label{eq:spearman}
\end{equation}
with $d_i$ the rank difference of item $i$. We also report top-1 agreement, the fraction of fields where
the reader's most-similar variant is the encoder's nearest neighbor,
\begin{equation}
\mathrm{Top\text{-}1} \;=\; \frac{1}{M}\sum_{k=1}^{M}
\mathbb{1}\!\left[\underset{j}{\arg\max}\; s^{\mathrm{rdr}}_{k,j}
\;=\; \underset{j}{\arg\min}\; D^{\mathrm{enc}}_{k,j}\right].
\label{eq:top1}
\end{equation}
The cross-cohort ratio and the stain-shift ratio reported below are Eq.~\eqref{eq:ratio} instantiated
with the in-house-versus-TCGA distance and with a same-tile graded stain shift as the numerator; a value
above one means the shift is detectable above the encoder's floor.

\section{Results}

\subsection{Raw distance is not comparable across encoders}

For the identical within-cohort comparison the raw distance ranged from 5.9 to 175.2 across the six
encoders (Table~\ref{tab:floor}), about a factor of thirty. The ordering does not follow embedding
dimension: the 1,280-dimensional Virchow2 gave the largest value and the 2,048-dimensional Inception-v3
one of the smallest. The spread reflects the magnitude and variance of each feature space, not its
dimensionality, so a raw distance cannot be interpreted or compared without naming the encoder.

\begin{table}[t]
\centering
\caption{Intrinsic distance floor (reference half versus reference half), matched $N=25{,}000$.}
\begin{tabular}{@{}llc@{}}
\toprule
Encoder & Dim & Within-cohort FD floor \\
\midrule
Phikon-v2 & 1024 & 5.9 \\
Inception-v3 & 2048 & 8.5 \\
CONCH & 512 & 14.7 \\
UNI2-h & 1536 & 35.9 \\
Prov-GigaPath & 1536 & 123.7 \\
Virchow2 & 1280 & 175.2 \\
\bottomrule
\end{tabular}
\label{tab:floor}
\end{table}

\subsection{A ratio normalization restores comparability}

\begin{figure}[t]
\centering
\includegraphics[width=\textwidth]{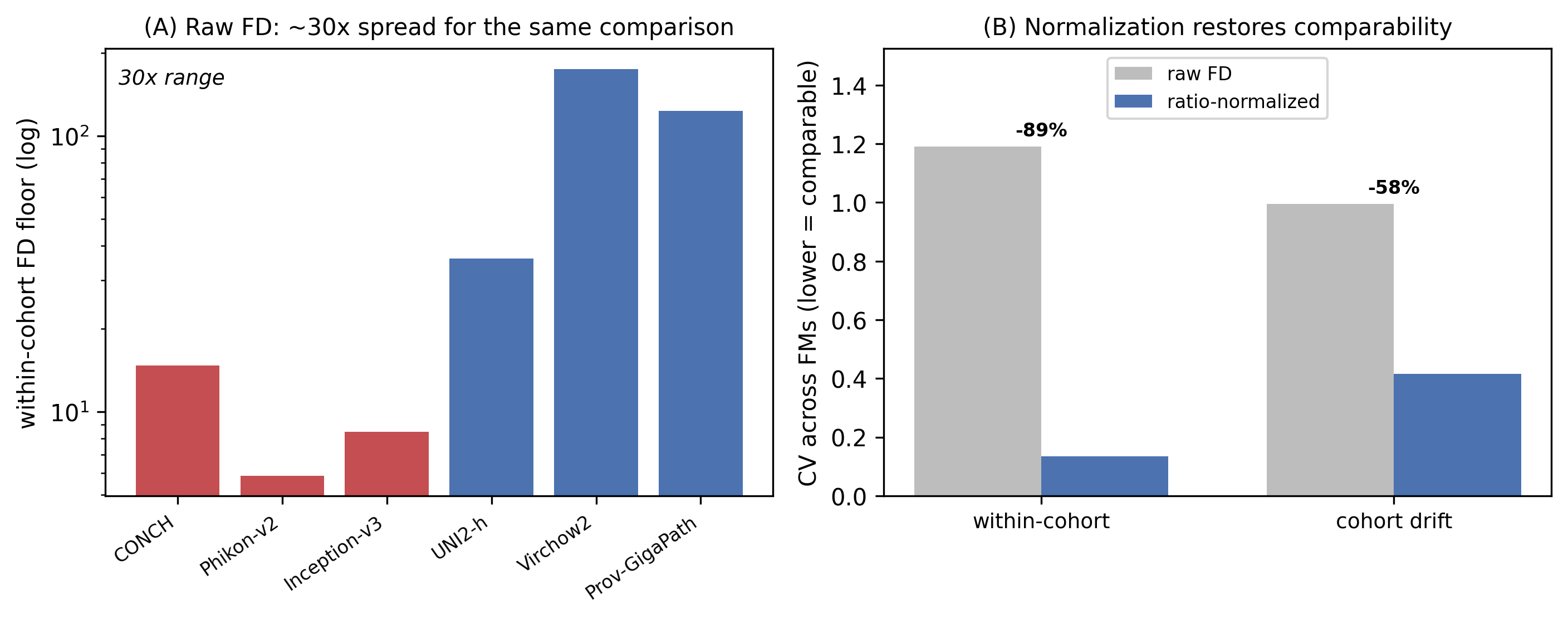}
\caption{(A) The within-cohort distance floor spans about thirty-fold across encoders for the same
comparison; bar color marks the two groups identified later. (B) The ratio normalization sharply reduces
the across-encoder coefficient of variation on two independent comparisons.}
\label{fig:norm}
\end{figure}

Dividing each distance by the encoder's own floor reduced the coefficient of variation across encoders
from 1.19 to 0.13 within cohort (about 89\%) and from 0.99 to 0.42 across cohorts (about 58\%)
(Figure~\ref{fig:norm}). Subtracting the floor removed only 2 to 3\%, because subtraction leaves each
value on the encoder's own scale. The ratio is therefore the normalization we use; its logarithm behaves
similarly.

\subsection{Two groups of encoders}

\begin{figure}[t]
\centering
\includegraphics[width=\textwidth]{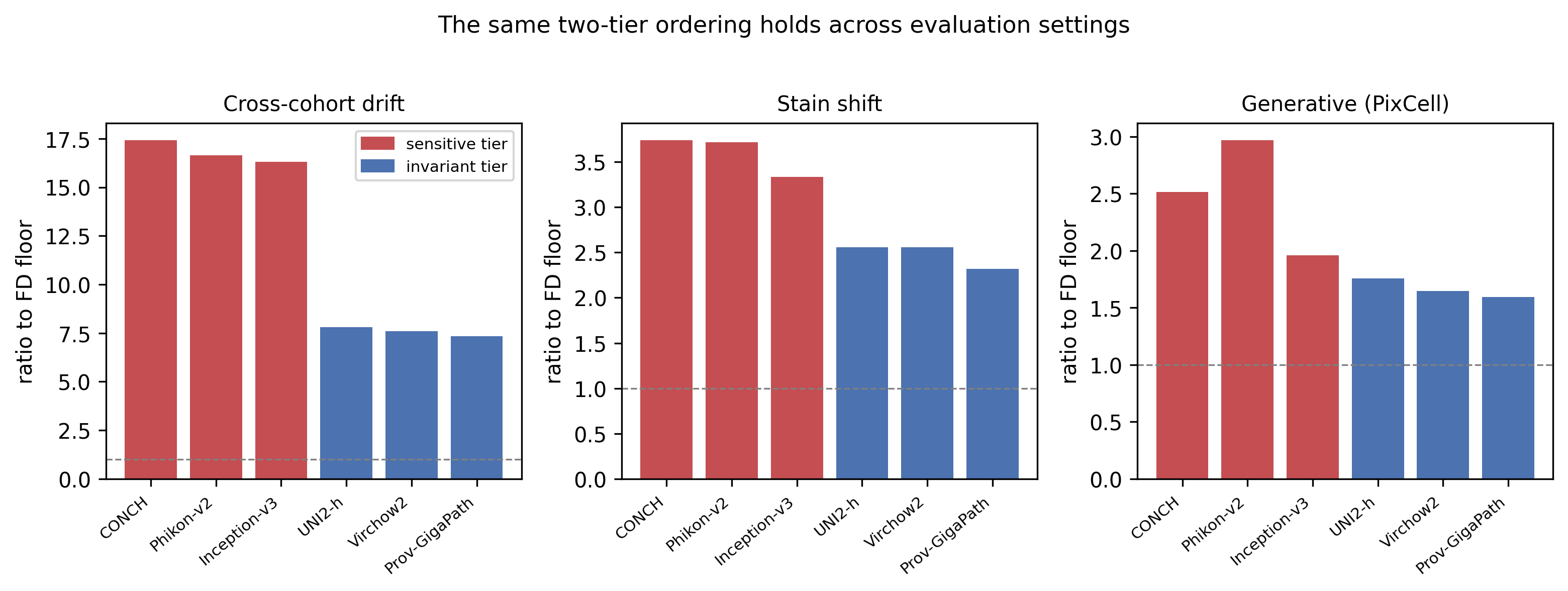}
\caption{The same two-group ordering of encoders appears for cross-cohort drift, stain shift and
generative evaluation. Bars are colored by group.}
\label{fig:twotier}
\end{figure}

All six encoders detected the in-house versus TCGA difference (ratio above one). After normalization they
fell into two groups (Table~\ref{tab:sensitivity}, Figure~\ref{fig:twotier}): CONCH, Phikon-v2 and
Inception-v3 reported a shift of 16 to 17 times the floor, while UNI2-h, Virchow2 and Prov-GigaPath
reported 7 to 8 times. The second group consists of the larger DINOv2 encoders trained on hundreds of
millions to billions of mostly proprietary tiles; they are comparatively invariant to cross-cohort
differences and so understate drift relative to the first group. The grouping does not reduce to the
training objective alone, since Phikon-v2 is also DINOv2-trained yet sits with the sensitive group;
feature scale, cohort breadth and augmentation strength all contribute.

\begin{table}[t]
\centering
\caption{Normalized sensitivity to cross-cohort drift and stain shift, and feature-level rotation
stability (cosine between a tile and its 90-degree rotation; 1.0 = stable).}
\begin{tabular}{@{}lccc@{}}
\toprule
Encoder & Cross-cohort ratio & Stain-shift ratio & Rotation cosine \\
\midrule
CONCH & 17.4 & 3.74 & 0.977 \\
Phikon-v2 & 16.6 & 3.72 & 0.719 \\
Inception-v3 & 16.3 & 3.34 & 0.846 \\
UNI2-h & 7.8 & 2.56 & 0.732 \\
Virchow2 & 7.6 & 2.56 & 0.949 \\
Prov-GigaPath & 7.3 & 2.32 & 0.917 \\
\bottomrule
\end{tabular}
\label{tab:sensitivity}
\end{table}

\subsection{Learned invariances}

\begin{figure}[tbp]
\centering
\includegraphics[width=0.78\textwidth]{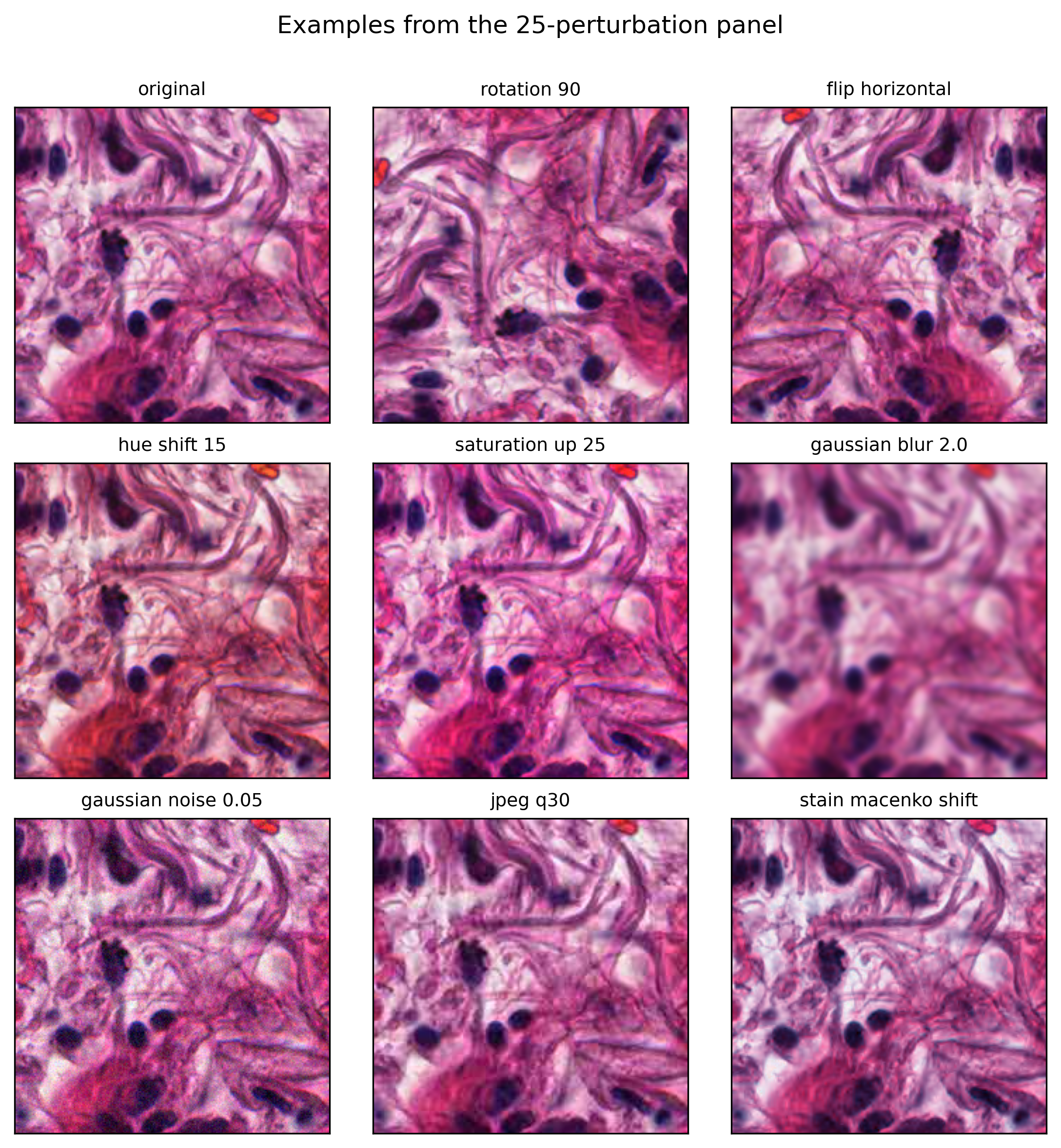}
\caption{Representative members of the 25-perturbation panel applied to one tile: geometric, photometric,
blur, noise, compression and stain shifts.}
\label{fig:perturb}
\end{figure}

The perturbation panel gave each encoder a distinct profile (Figure~\ref{fig:perturb}). Geometric
stability did not track the sensitive/invariant grouping. The feature-level cosine between a tile and its
90-degree rotation ranged from 0.98 (CONCH) and 0.95 (Virchow2) down to 0.73 (UNI2-h) and 0.72
(Phikon-v2), with Prov-GigaPath (0.92) and Inception-v3 (0.85) in between (Table~\ref{tab:sensitivity}).
Rotation stability is therefore a separate axis: Virchow2 and UNI2-h fall in the same invariant cohort
and stain group yet differ sharply here. No encoder was fully invariant, and the distance amplified the
residual: a cosine of 0.92 still produced a large rotation distance, so the distance is a stricter probe
of invariance than cosine similarity. Table~\ref{tab:perturb} gives the per-encoder sensitivity for a
subset, normalized within each encoder so values are comparable across encoders. Noise produced the
largest distance for every encoder.

\subsection{Stain normalization}

Sensitivity to the raw stain difference followed the grouping of Table~\ref{tab:sensitivity}, with the
sensitive encoders reporting larger shifts. Applying Macenko or Reinhard normalization changed the
distance only marginally and sometimes increased it, because the TCGA to in-house difference is one of
tissue content and stain panel, not only color.

\begin{table}[t]
\centering
\caption{Within-encoder normalized perturbation sensitivity (0 = invariant, 1 = that encoder's most
disruptive perturbation). CONCH is near-invariant to rotation and flip; every encoder is most sensitive
to noise.}
\begin{tabular}{@{}lcccccc@{}}
\toprule
Encoder & rot 90 & flip horiz. & hue shift & blur 2.0 & jpeg q30 & salt-pepper \\
\midrule
CONCH & 0.01 & 0.00 & 0.01 & 0.22 & 0.24 & 1.00 \\
Phikon-v2 & 0.36 & 0.01 & 0.01 & 0.03 & 0.69 & 1.00 \\
Inception-v3 & 0.34 & 0.11 & 0.03 & 0.42 & 0.25 & 1.00 \\
UNI2-h & 0.51 & 0.09 & 0.02 & 0.06 & 0.73 & 1.00 \\
Virchow2 & 0.34 & 0.03 & 0.02 & 0.03 & 0.47 & 1.00 \\
Prov-GigaPath & 0.41 & 0.19 & 0.11 & 0.13 & 1.00 & 0.74 \\
\bottomrule
\end{tabular}
\label{tab:perturb}
\end{table}

\subsection{Compression}

\begin{figure}[tbp]
\centering
\setlength{\tabcolsep}{2.5pt}
\begin{tabular}{ccccc}
\small\textbf{Original} & \small\textbf{JPEG} & \small\textbf{JPEG 2000} & \small\textbf{JPEG XL} & \small\textbf{TuroCompress}\\[2pt]
\includegraphics[width=0.185\textwidth]{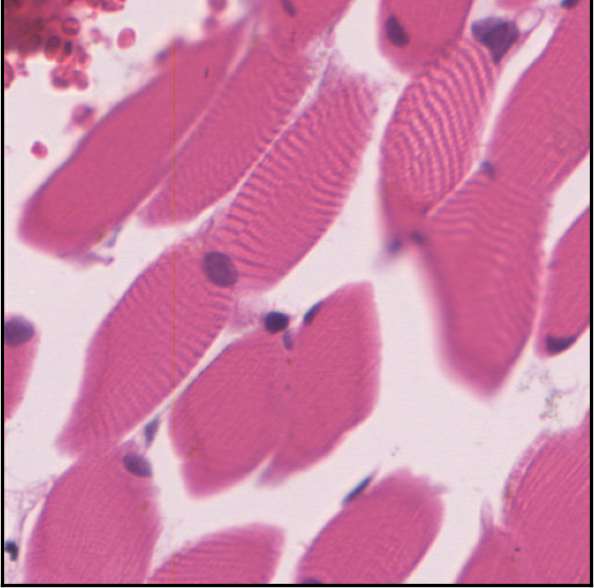} &
\includegraphics[width=0.185\textwidth]{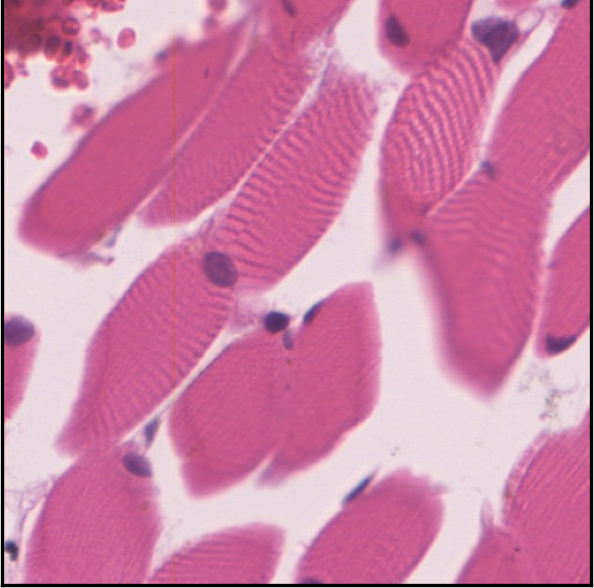} &
\includegraphics[width=0.185\textwidth]{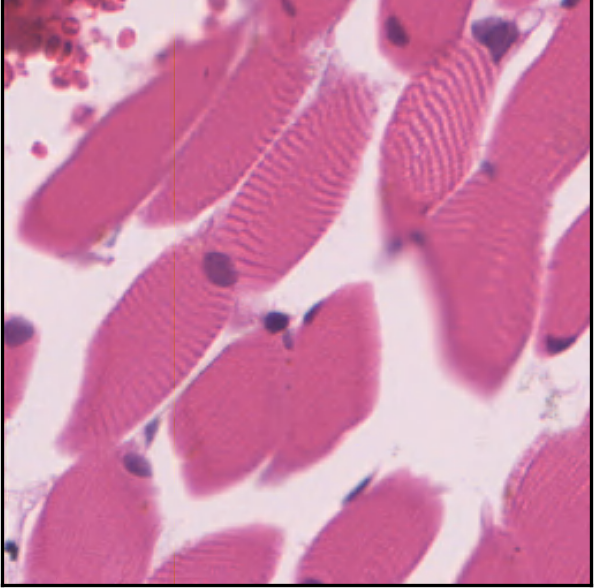} &
\includegraphics[width=0.185\textwidth]{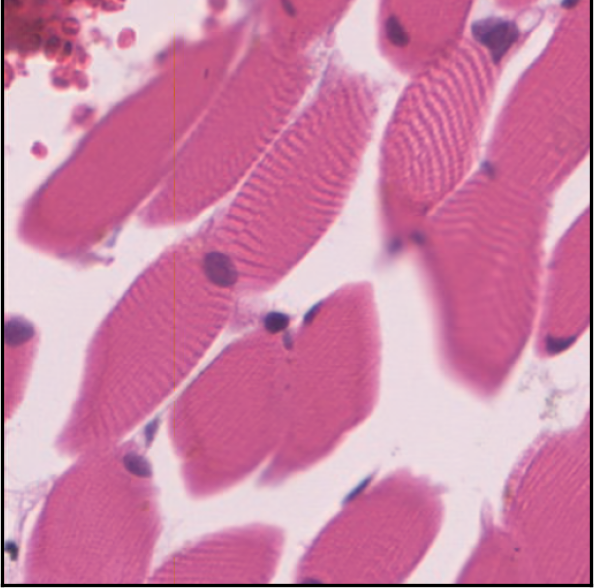} &
\includegraphics[width=0.185\textwidth]{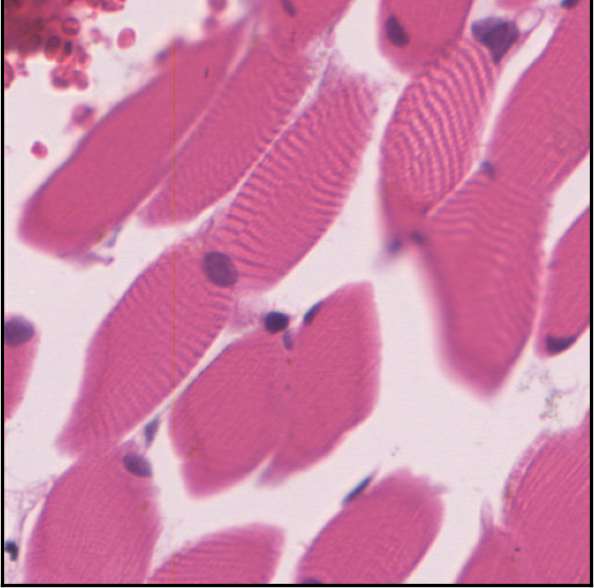}\\[1pt]
{\footnotesize 347 kB (PNG)} & {\footnotesize\shortstack{49 kB\\SSIM 0.936}} &
{\footnotesize\shortstack{20 kB\\SSIM 0.918}} & {\footnotesize\shortstack{18 kB\\SSIM 0.929}} &
{\footnotesize\shortstack{6.5 kB\\SSIM $>$0.95}}\\[5pt]
\includegraphics[width=0.185\textwidth]{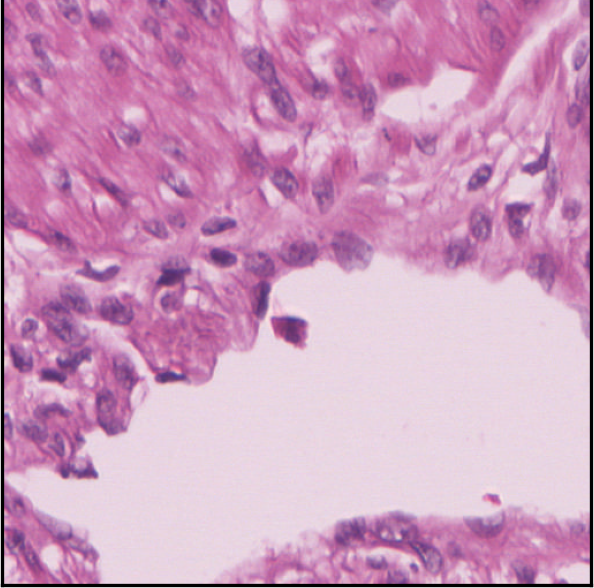} &
\includegraphics[width=0.185\textwidth]{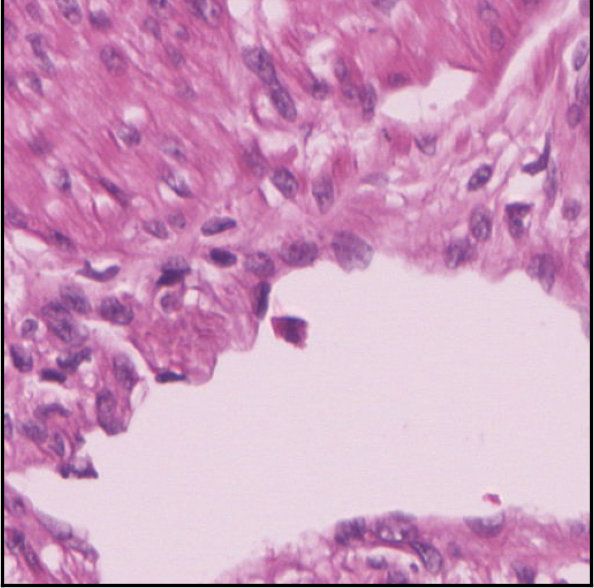} &
\includegraphics[width=0.185\textwidth]{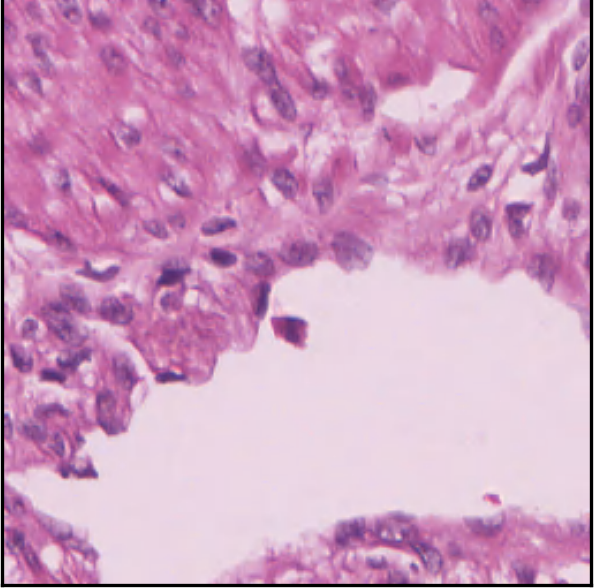} &
\includegraphics[width=0.185\textwidth]{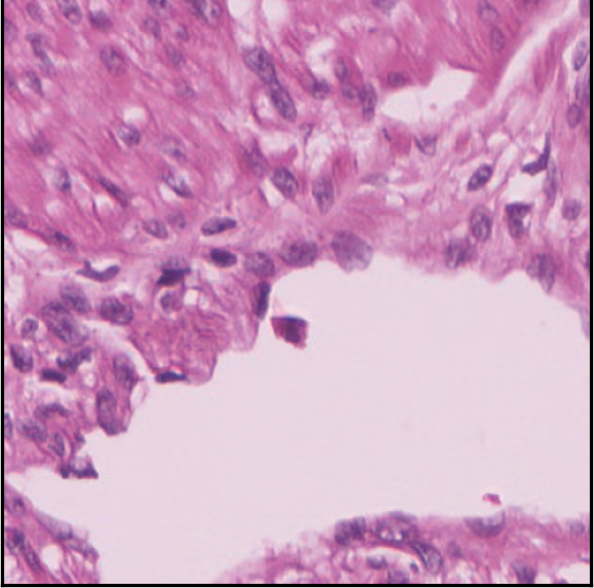} &
\includegraphics[width=0.185\textwidth]{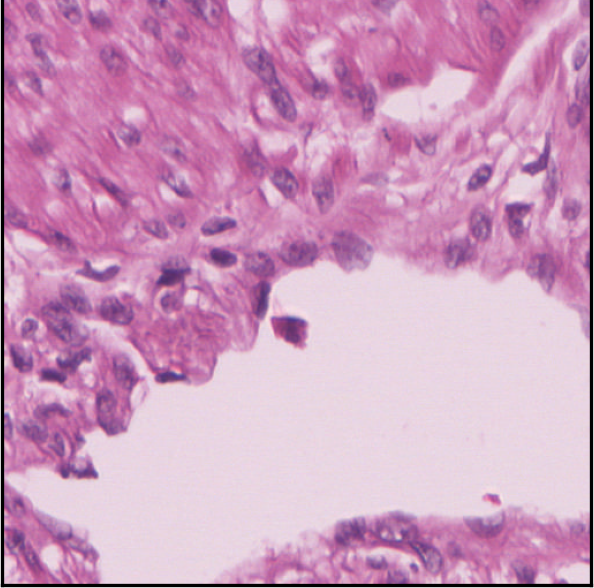}\\[1pt]
{\footnotesize 369 kB (PNG)} & {\footnotesize\shortstack{49 kB\\SSIM 0.926}} &
{\footnotesize\shortstack{19 kB\\SSIM 0.909}} & {\footnotesize\shortstack{17 kB\\SSIM 0.917}} &
{\footnotesize\shortstack{6.5 kB\\SSIM $>$0.95}}\\
\end{tabular}
\caption{Two already-compressed H\&E tiles (lossless-PNG sizes shown) re-compressed at each codec's
operating point. All preserve diagnostic structure; TuroCompress does so at the smallest size and the
highest fidelity.}
\label{fig:compression}
\end{figure}

The in-house tiles are already-compressed histology crops; as lossless PNG they average 355 kB per tile
(14.5 MB for a 40-tile sample). At its operating point TuroCompress reaches about 6.5 kB per tile, a
55-fold reduction, while holding SSIM above 0.95, whereas JPEG, JPEG 2000 and JPEG XL need roughly 50 kB,
20 kB and 15 kB to reach comparable quality, a 7 to 24-fold reduction (Table~\ref{tab:compression};
TuroCompress figures are from the production Turocrates.ai pipeline). Visually all four preserve
diagnostic structure at these operating points, with TuroCompress holding the smallest size at the
highest fidelity (Figure~\ref{fig:compression}). Encoder distance tracked SSIM closely across codecs
(Spearman 0.93 to 0.98), so for compression artifacts the encoder distance adds little beyond pixel
metrics; at matched size TuroCompress gave the lowest encoder distance among the codecs.

\begin{table}[t]
\centering
\caption{Compression of already-compressed H\&E tiles at each codec's operating point. Reduction is
relative to the lossless-PNG size. JPEG, JPEG 2000 and JPEG XL are measured here (mean over 40 tiles);
TuroCompress is the production Turocrates.ai pipeline. TuroCompress reaches the largest reduction at the
highest quality.}
\begin{tabular}{@{}lccc@{}}
\toprule
Codec & Avg size (kB) & Reduction & SSIM \\
\midrule
Original (PNG) & 355 & 1$\times$ & n/a \\
JPEG & 50 & 7$\times$ & 0.93 \\
JPEG 2000 & 20 & 18$\times$ & 0.92 \\
JPEG XL & 15 & 24$\times$ & 0.94 \\
TuroCompress & 6.5 & \textbf{55}$\times$ & \textbf{$>$0.95} \\
\bottomrule
\end{tabular}
\label{tab:compression}
\end{table}

\subsection{Generative evaluation depends on the encoder}

PixCell-256 and CytoSyn were scored against the same in-house reference (Table~\ref{tab:generative}).
CytoSyn obtained a lower distance ratio for five of six encoders and substantially higher recall (0.09 to
0.17 versus near zero), consistent with broader coverage of the real distribution; PixCell's near-zero
recall is consistent with narrower coverage in these feature spaces. The two generators were ranked
differently depending on the encoder: under CONCH the gap was large in PixCell's disfavor (1.42 versus
2.52), whereas under Inception-v3 and Phikon-v2 they were close. A generative-quality conclusion in
pathology therefore depends on the encoder used to compute it. Both ratios stayed above one, reflecting a
domain gap between either generator's training distribution and the in-house cohort; PixCell was used
conditionally and CytoSyn unconditionally, an asymmetry to note.

\begin{table}[t]
\centering
\caption{PixCell-256 versus CytoSyn (lower distance ratio = closer to the real reference).}
\begin{tabular}{@{}lcccc@{}}
\toprule
Encoder & PixCell ratio & CytoSyn ratio & PixCell recall & CytoSyn recall \\
\midrule
Inception-v3 & 1.96 & 2.14 & 0.072 & 0.170 \\
Phikon-v2 & 2.97 & 2.89 & 0.017 & 0.002 \\
CONCH & 2.52 & 1.42 & 0.000 & 0.016 \\
UNI2-h & 1.76 & 1.52 & 0.001 & 0.010 \\
Virchow2 & 1.65 & 1.49 & 0.001 & 0.094 \\
Prov-GigaPath & 1.59 & 1.44 & 0.014 & 0.165 \\
\bottomrule
\end{tabular}
\label{tab:generative}
\end{table}

\subsection{Slide-level distance}

A patch-level Fr\'echet distance pools all tiles and is blind to how tiles are arranged within slides. We
tested whether a slide-level distance recovers that information using two encoders that aggregate patch
features into one vector per slide: the Prov-GigaPath LongNet encoder and CHIEF. CTransPath features feed
CHIEF and Prov-GigaPath tile features feed the LongNet encoder. The in-house tiles carry no stored
coordinates, so a common grid layout is used for both cohorts to keep the comparison fair.

\textit{Slide-level distance sees per-slide composition that patch distance misses.} We built two TCGA
cohorts with identical pooled-tile composition but opposite per-slide composition: in one, each slide is
homogeneous (its tiles come from a single tissue cluster); in the other, each slide mixes both clusters
equally. Because the pooled tiles match, the patch-level distance stays at the floor (CHIEF 0.026,
Prov-GigaPath 2.11). The CHIEF slide-level distance instead rises to 8.36, about 320 times the
patch-level value (Table~\ref{tab:slide}), recovering the per-slide structure the patch distance cannot
see. The Prov-GigaPath LongNet encoder, whose positional model expects true tile coordinates that the
remixed pseudo-slides do not provide, shows no such amplification (ratio near one), so the effect is a
property of the aggregator rather than of slide-level distance in general.

\textit{Slide-level cohort drift.} Between the in-house and TCGA cohorts the slide-level distance
separates the two strongly under both encoders, with a drift-to-floor ratio of 332 for CHIEF and 42 for
Prov-GigaPath (the within-cohort floor splits the in-house slides in half). Cohort drift is therefore at
least as pronounced at the slide level as at the patch level.

\begin{table}[t]
\centering
\caption{Slide-level distance. Patch-versus-slide uses two TCGA cohorts with matched pooled-tile but
opposite per-slide composition; the slide distance amplifies the per-slide signal for the
attention-pooling encoder only. Cohort drift is in-house versus TCGA as a ratio to the within-cohort
floor.}
\begin{tabular}{@{}lcc@{}}
\toprule
 & CHIEF (attention pool) & Prov-GigaPath (LongNet) \\
\midrule
Patch-level FD (matched cohorts) & 0.026 & 2.11 \\
Slide-level FD (matched cohorts) & \textbf{8.36} & 2.11 \\
Slide / patch ratio & \textbf{322} & 1.0 \\
Cohort-drift ratio (in-house vs TCGA) & 332 & 42 \\
\bottomrule
\end{tabular}
\label{tab:slide}
\end{table}

\subsection{Pathologist concordance}

To test whether the Fr\'echet-distance signal reflects how a pathologist perceives image difference, two
board-certified pathologists reviewed de-identified H\&E fields under a blinded protocol (no encoder names
or condition labels shown) and, for each field, ordered a set of variants by their similarity to a
reference and rated selected pairs on an absolute similarity scale (1 = identical to 5 = very different).

\textit{Fr\'echet distance tracks the expert eye.} Across stain variation, focus degradation and codec
comparisons, the perceived similarity ordering agreed closely with the ordering induced by
foundation-model feature distance (Spearman $\rho = 0.96$, $p < 0.05$; the most-similar variant chosen by
the reader matched the encoder's nearest variant in 86\% of fields). The direction in which images were
ranked and the direction in which the Fr\'echet distance grew coincided: as compression rate and other
degradations increased, perceived difference and Fr\'echet distortion rose together. Reader-perceived and
distance-based assessments of image difference are therefore in line (Table~\ref{tab:concordance}).

\textit{No single encoder was perceptually privileged.} This concordance held across both encoder families
identified above, the sensitivity-preserving encoders (e.g.\ Phikon-v2) and the augmentation-invariant
DINOv2-style encoders (e.g.\ Virchow2), as well as the Inception-v3 benchmark. The two-tier divergence in
normalized Fr\'echet distance reported above is therefore a property of each encoder's feature geometry
and sensitivity scale, not a sign that either family contradicts how a pathologist perceives tissue
similarity: both tiers agree on the direction of difference and differ only on its magnitude in feature
space.

\textit{TuroCompress preserves diagnostic appearance.} When the reconstructed fields were compared
directly against their originals on the absolute scale, every pair was rated at the most-similar end
(median rating 1 of 5, "identical", with no pair rated worse than "somewhat similar"); the reconstructions
were judged diagnostically indistinguishable from the source. This is consistent with the objective
compression result: at its operating point TuroCompress attains the lowest reconstruction distance among
the codecs evaluated while a pathologist judges its output equivalent to the original.

\begin{table}[t]
\centering
\caption{Pathologist concordance: perceived similarity ordering versus feature-distance ordering over the
ranking fields. Concordance is uniformly high across encoder families.}
\begin{tabular}{@{}lc@{}}
\toprule
Feature extractor & Spearman $\rho$ vs reader ($p<0.05$) \\
\midrule
Phikon-v2 (sensitive tier) & 0.96 \\
Virchow2 (invariant tier) & 0.96 \\
Inception-v3 (benchmark) & 0.95 \\
\bottomrule
\end{tabular}
\label{tab:concordance}
\end{table}

\section{Discussion}

A raw Fr\'echet distance computed with a pathology encoder is a property of the feature space as much as
of the images, and is not comparable across encoders. The ratio to the encoder's own floor removes most
of this dependence and yields numbers that can be compared and combined. Once comparable, the encoders
divide into sensitive and invariant groups, and the division is consistent across cohort drift, stain
shift and generative evaluation. It does not reduce to the self-supervision objective, since a DINOv2
encoder appears in each group; feature scale, training-cohort breadth and augmentation strength jointly
determine where an encoder falls.

The practical guidance is to match the encoder to the question. To detect scanner or cohort drift or to
flag stain differences, a sensitive encoder is preferable. To measure a quantity that should be invariant
to such nuisances, a strongly augmented DINOv2 encoder is appropriate, accepting that it will understate
nuisance shift. For generative evaluation the encoder can change the ranking of models, so we recommend
reporting normalized distance under more than one encoder and reporting recall alongside it.

\section{Conclusion}

The Fr\'echet distance is useful in computational pathology only when its dependence on the feature
extractor is acknowledged and removed. A ratio to the encoder's own within-cohort floor restores
comparability, after which extractors separate into consistent groups whose behavior should guide
extractor choice. We release the protocol, the per-encoder perturbation panel and the feature extracts to
support reproducible evaluation.


\begin{thebibliography}{99}

\bibitem{heusel2017} Heusel M, Ramsauer H, Unterthiner T, Nessler B, Hochreiter S. GANs trained by a two
time-scale update rule converge to a local Nash equilibrium. NeurIPS, 2017.

\bibitem{dowson1982} Dowson DC, Landau BV. The Fr\'echet distance between multivariate normal
distributions. Journal of Multivariate Analysis, 1982;12(3):450--455.

\bibitem{szegedy2016} Szegedy C, Vanhoucke V, Ioffe S, Shlens J, Wojna Z. Rethinking the Inception
architecture for computer vision. CVPR, 2016, pp. 2818--2826.

\bibitem{chong2020} Chong MJ, Forsyth D. Effectively unbiased FID and Inception Score and where to find
them. CVPR, 2020. arXiv:1911.07023.

\bibitem{oquab2024dinov2} Oquab M, Darcet T, Moutakanni T, et al. DINOv2: learning robust visual features
without supervision. TMLR, 2024. arXiv:2304.07193.

\bibitem{chen2024uni} Chen RJ, Ding T, Lu MY, et al. Towards a general-purpose foundation model for
computational pathology. Nature Medicine, 2024;30:850--862.

\bibitem{vorontsov2024virchow} Vorontsov E, Bozkurt A, Casson A, et al. A foundation model for
clinical-grade computational pathology and rare cancers detection. Nature Medicine, 2024;30:2924--2935.

\bibitem{zimmermann2024virchow2} Zimmermann E, Vorontsov E, Viret J, et al. Virchow2: scaling
self-supervised mixed magnification models in pathology. arXiv:2408.00738, 2024.

\bibitem{filiot2023} Filiot A, Ghermi R, Olivier A, et al. Scaling self-supervised learning for
histopathology with masked image modeling. medRxiv 2023.07.21.23292757, 2023.

\bibitem{filiot2024phikon} Filiot A, Jacob P, Mac Kain A, Saillard C. Phikon-v2: a large and public
feature extractor for biomarker prediction. arXiv:2409.09173, 2024.

\bibitem{lu2024conch} Lu MY, Chen B, Williamson DFK, et al. A visual-language foundation model for
computational pathology. Nature Medicine, 2024;30:863--874.

\bibitem{xu2024gigapath} Xu H, Usuyama N, Bagga J, et al. A whole-slide foundation model for digital
pathology from real-world data. Nature, 2024;630:181--188.

\bibitem{kynkaanniemi2019} Kynk\"a\"anniemi T, Karras T, Laine S, Lehtinen J, Aila T. Improved precision
and recall metric for assessing generative models. NeurIPS, 2019. arXiv:1904.06991.

\bibitem{jiralerspong2023} Jiralerspong M, Bose J, Gemp I, et al. Feature Likelihood Divergence:
evaluating the generalization of generative models using samples. NeurIPS, 2023. arXiv:2302.04440.

\bibitem{grossman2016} Grossman RL, Heath AP, Ferretti V, et al. Toward a shared vision for cancer genomic
data (NCI Genomic Data Commons). New England Journal of Medicine, 2016;375:1109--1112.

\bibitem{yellapragada2025pixcell} Yellapragada S, Graikos A, Triaridis K, et al. PixCell: a generative
foundation model for digital histopathology images. arXiv:2506.05127, 2025.

\bibitem{ma2024sit} Ma N, Goldstein M, Albergo MS, Boffi NM, Vanden-Eijnden E, Xie S. SiT: exploring flow
and diffusion-based generative models with scalable interpolant transformers. ECCV, 2024. arXiv:2401.08740.

\bibitem{leng2025repae} Leng X, Singh J, Hou Y, Xing Z, Xie S, Zheng L. REPA-E: unlocking VAE for
end-to-end tuning with latent diffusion transformers. arXiv:2504.10483, 2025.

\bibitem{macenko2009} Macenko M, Niethammer M, Marron JS, et al. A method for normalizing histology slides
for quantitative analysis. IEEE ISBI, 2009, pp. 1107--1110.

\bibitem{reinhard2001} Reinhard E, Ashikhmin M, Gooch B, Shirley P. Color transfer between images. IEEE
Computer Graphics and Applications, 2001;21(5):34--41.

\bibitem{vahadane2016} Vahadane A, Peng T, Sethi A, et al. Structure-preserving color normalization and
sparse stain separation for histological images. IEEE TMI, 2016;35(8):1962--1971.

\bibitem{wang2004} Wang Z, Bovik AC, Sheikh HR, Simoncelli EP. Image quality assessment: from error
visibility to structural similarity. IEEE TIP, 2004;13(4):600--612.

\bibitem{zhang2018} Zhang R, Isola P, Efros AA, Shechtman E, Wang O. The unreasonable effectiveness of
deep features as a perceptual metric. CVPR, 2018, pp. 586--595.

\end{thebibliography}
\end{document}